\documentclass[sigconf,nonacm,9pt]{acmart}
\AtBeginDocument{%
  \providecommand\BibTeX{{%
    Bib\TeX}}}

\usepackage{amsmath,amssymb,amsfonts}
\usepackage{graphicx}
\usepackage{textcomp}
\usepackage{booktabs} 
\usepackage{multirow}
\usepackage{setspace}
\usepackage{hyperref}
\usepackage{url}
\usepackage{nicefrac}
\usepackage{microtype}
\usepackage{xcolor}
\usepackage{balance}
\usepackage{stfloats}
\usepackage{tikz}
\usetikzlibrary{positioning,arrows.meta,shapes.geometric,calc}
\usepackage{placeins}

\usepackage{orcidlink}
\usepackage{comment}
\usepackage{braket}
\def\BibTeX{{\rm B\kern-.05em{\sc i\kern-.025em b}\kern-.08em
    T\kern-.1667em\lower.7ex\hbox{E}\kern-.125emX}}
\usepackage{pifont}
\usepackage{amsmath,amssymb,amsfonts}
\usepackage{graphicx}
\usepackage{textcomp}
\usepackage{amsmath}
\usepackage{booktabs} 
\usepackage{color}
\usepackage{multirow}
\usepackage{colortbl, xcolor}
\usepackage{hyperref}       % hyperlinks
\usepackage{url}            % simple URL typesetting
\usepackage{booktabs}       % professional-quality tables
\usepackage{amsfonts}       % blackboard math symbols
\usepackage{nicefrac}       % compact symbols for 1/2, etc.
\usepackage{microtype}      % microtypography
\usepackage{xcolor}         % colors

\usepackage{booktabs}
\usepackage{siunitx}

\usepackage{amsmath}
\usepackage{tikz}

\usepackage{orcidlink}
\usepackage{comment}
\usepackage{amssymb}
\usepackage{braket}
\usepackage{xcolor}
\usepackage{cleveref}
\usepackage{algorithm}
\usepackage{algpseudocode}
\usepackage{float} % for [H] placement (optional)
\usepackage{enumitem}
\setlist[itemize]{leftmargin=*}%
\setlist[enumerate]{leftmargin=*}%
\usepackage[font=scriptsize,skip=0pt]{caption}
\begin{document}

\title{Late Breaking Results: Hardware-Efficient Quantum Reservoir Computing via Quantized Readout}

\author{
Param Pathak$^{1,*}$, 
Mansi Od$^{2,*}$, 
Nouhaila Innan$^{3,4,*, \ddagger}$, 
Muhammad Shafique$^{3,4}$\\
\small $^1$\textit{QuantumAI Lab}, \textit{Fractal Analytics}, Mumbai, Maharashtra, India\\
\small $^2$University of Greenwich, London, UK\\
\small $^3$eBRAIN Lab, Division of Engineering, New York University Abu Dhabi (NYUAD), Abu Dhabi, UAE\\
$^4$Center for Quantum and Topological Systems (CQTS), NYUAD Research Institute, NYUAD, Abu Dhabi, UAE\\
\vspace{-2pt}
}
\vspace{-3pt}

% =========================
\begin{abstract}
\vspace{-2pt}
Due to rising electricity demand, accurate short-term load forecasting is increasingly important for grid stability and efficient energy management, particularly in resource-constrained edge settings. We present a hardware-efficient Quantum Reservoir Computing (QRC) framework based on a fixed, untrained quantum circuit with Chebyshev feature encoding, brickwork entanglement, and single- and two-qubit Pauli measurements, avoiding quantum backpropagation entirely. Using the Tetouan City Power Consumption dataset, we examine the effect of post-training fixed-point quantization on the classical readout layer, with the reservoir architecture selected through a genetic search over 18 candidate configurations. Under finite-shot evaluation, 8-bit and 6-bit quantization maintain forecasting accuracy within 1\% of the FP32 baseline while reducing readout memory by 75\% and 81\%, respectively. These results suggest that quantized readout can improve the hardware efficiency and deployment practicality of QRC for memory-constrained energy forecasting.
\vspace{-2pt}
\end{abstract}

% \keywords{\vspace{-2pt}Quantum Reservoir Computing, Quantized Readout, Energy Forecasting, Edge Deployment, Finite-Shot Evaluation}

\maketitle

\renewcommand{\shortauthors}{Pathak, Od, Innan, and Shafique}
\begin{spacing}{0.96}

\vspace{-4pt}
%\vspace{-8pt}
\section{Introduction}
\vspace{-1pt}
\begingroup
\renewcommand{\thefootnote}{}
\footnotetext[0]{$^*$\textbf{These authors contributed equally to this work.\\$\ddagger$ nouhaila.innan@nyu.edu}}
\addtocounter{footnote}{-1}
\endgroup

Global electricity consumption is rising, and grid operators increasingly rely on short-term forecasting to balance supply and demand in real time \cite{asiri2024short}. Accurate zone-level forecasting on embedded hardware places strict constraints on model size, memory, and compute cost, making efficient and deployable forecasting models increasingly important.
Recurrent neural networks and transformer-based models have shown strong performance on energy forecasting benchmarks \cite{moustati2025unveiling}, but their parameter counts and computational cost can limit deployment on resource-constrained edge devices. Reservoir Computing offers a simpler alternative: a fixed, randomly initialized dynamical system maps inputs into a rich feature space, while only a small output layer is trained. Classical echo state networks follow this design, but their capacity scales with the size of the physical reservoir. Quantum Reservoir Computing (QRC) extends this idea by replacing the classical reservoir with a quantum circuit, where superposition and entanglement can generate expressive feature representations from relatively few qubits without requiring gradient-based quantum training \cite{fujii2017harnessing}.

Despite this potential, two practical challenges limit the deployment of QRC for edge forecasting. First, many prior studies focus on ideal noiseless simulation, whereas realistic execution must account for finite-shot measurement effects. Second, the effect of post-training quantization on QRC readout accuracy remains largely unexplored, leaving the compression-accuracy tradeoff unclear for memory-constrained deployment settings \cite{abbas2024classical}.
To address these challenges, this paper makes the following contributions:
\begin{enumerate}
%\vspace{-2pt}
\item \textbf{Hardware-efficient QRC pipeline:}
We develop a QRC framework with Chebyshev encoding, brickwork entanglement, and Pauli-based measurements, with the reservoir architecture selected by genetic algorithm search over 18 candidates and no quantum training.

\item \textbf{Finite-shot evaluation:}
We evaluate the full pipeline under finite-shot settings (\texttt{shots}=512) across two random seeds, reporting mean and standard deviation to quantify the effect of measurement noise.

\item \textbf{Post-training quantization analysis:}
We quantize the classical readout to 8, 6, 4, 3, and 2 bits, and show that 8-bit and 6-bit maintain accuracy within 1\% of FP32 while reducing readout memory by 75\% and 81\%, respectively.
\end{enumerate}

\section{Methodology}
\begin{figure}[htpb]
    \centering
    \includegraphics[width=1\linewidth]{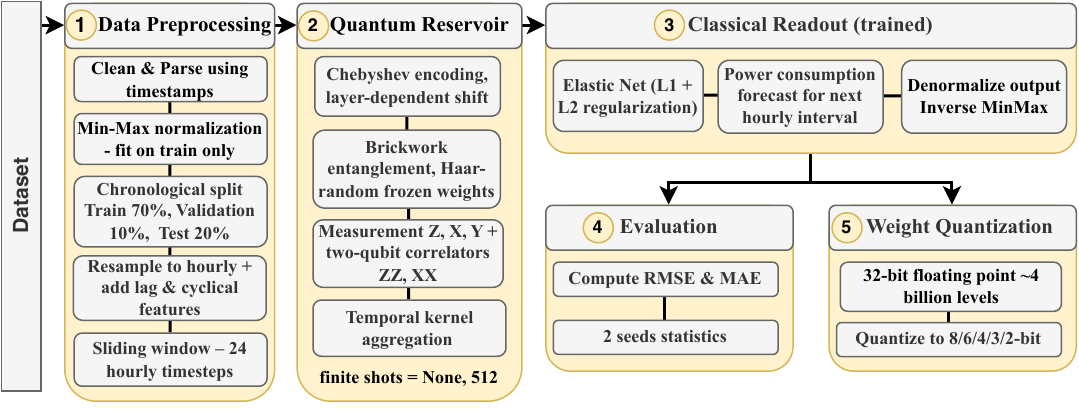}
    \caption{QRC pipeline: hourly-resampled Tetouan data flows through GA-optimized quantum reservoir extraction, temporal aggregation, and Elastic Net readout, followed by post-training fixed-point quantization for edge deployment.}
    \label{wf}
\end{figure}

\vspace{-1pt}
The proposed hardware-efficient QRC framework consists of five stages: data preprocessing, quantum reservoir feature extraction, classical readout training, evaluation, and post-training weight quantization for edge deployment, as illustrated in Fig.~\ref{wf}. We use the Tetouan City Power Consumption dataset \cite{power_consumption_of_tetouan_city_849}, which contains 52,416 samples collected at 10-minute intervals throughout 2017. The data are resampled to hourly resolution, yielding 8,736 samples with eleven input features, including temperature, humidity, wind speed, general and diffuse flows, cyclical time encodings, and lag features. All features are normalized to $[0,1]$ using Min-Max scaling fit only on the training set to avoid data leakage. A strict chronological split of 70\% for training, 10\% for validation, and 20\% for testing is then applied without shuffling to preserve temporal structure. A sliding window of $T=24$ hourly time steps is used to construct supervised samples, where each input window is mapped to the immediately following time step as the prediction target.

The QRC architecture is selected using a genetic algorithm over 18 candidate configurations, resulting in a final design with $N=7$ qubits and $L=4$ layers. The search space includes qubit count $\in \{5,6,7\}$, layer depth $\in \{3,4,5\}$, encoding strategy, coupling strength, and regularization ratio, using a population of 6 over 3 generations with tournament selection, crossover, and mutation. To reduce search cost, candidate architectures are evaluated on 20\% of the training set. Reservoir parameters are Haar-randomly initialized and then kept fixed throughout, avoiding gradient-based quantum training. Input features are encoded using Chebyshev rotations with layer-dependent shifts, followed by fixed brickwork entanglement layers. At each time step, we measure single-qubit Pauli observables $Z$, $X$, and $Y$, together with nearest-neighbor two-qubit correlators $ZZ$ and $XX$. These measurements are aggregated across the input window using exponential temporal kernels to form the final feature vector $\mathbf{r}$. A classical Elastic-Net readout with combined $\ell_1$ and $\ell_2$ regularization is then trained to predict power consumption.

Model performance is evaluated using RMSE and MAE under finite-shot settings of $\{\texttt{None}, 512\}$ and repeated across two random seeds, with results reported as mean $\pm$ standard deviation. Quantum circuit simulations are implemented in PennyLane and accelerated on an NVIDIA A100 GPU via Google Colab Pro+, while data preprocessing, architecture search, and readout training are performed separately on a local Apple M4 system with 32\,GB unified memory.

To improve deployment efficiency, we apply post-training fixed-point quantization to the trained readout parameters. For bit width $k \in \{8,6,4,3,2\}$, the quantized prediction is given by $ \scriptsize \hat{y}^{(k)} = ({w}^{(k)})^{\top}{r}$ $+ b^{(k)},$
where ${r}$ denotes the aggregated reservoir feature vector, and ${w}^{(k)}$ and $b^{(k)}$ are the quantized readout weights and bias obtained from the trained FP32 parameters using iterative refinement with optimal clipping. This enables direct comparison between full-precision and low-precision readout under identical reservoir features and finite-shot settings.
%\vspace{-10pt}
\vspace{-2pt}
\section{Results}
\begin{figure}[h]
    \centering
    \includegraphics[width=0.91\linewidth]{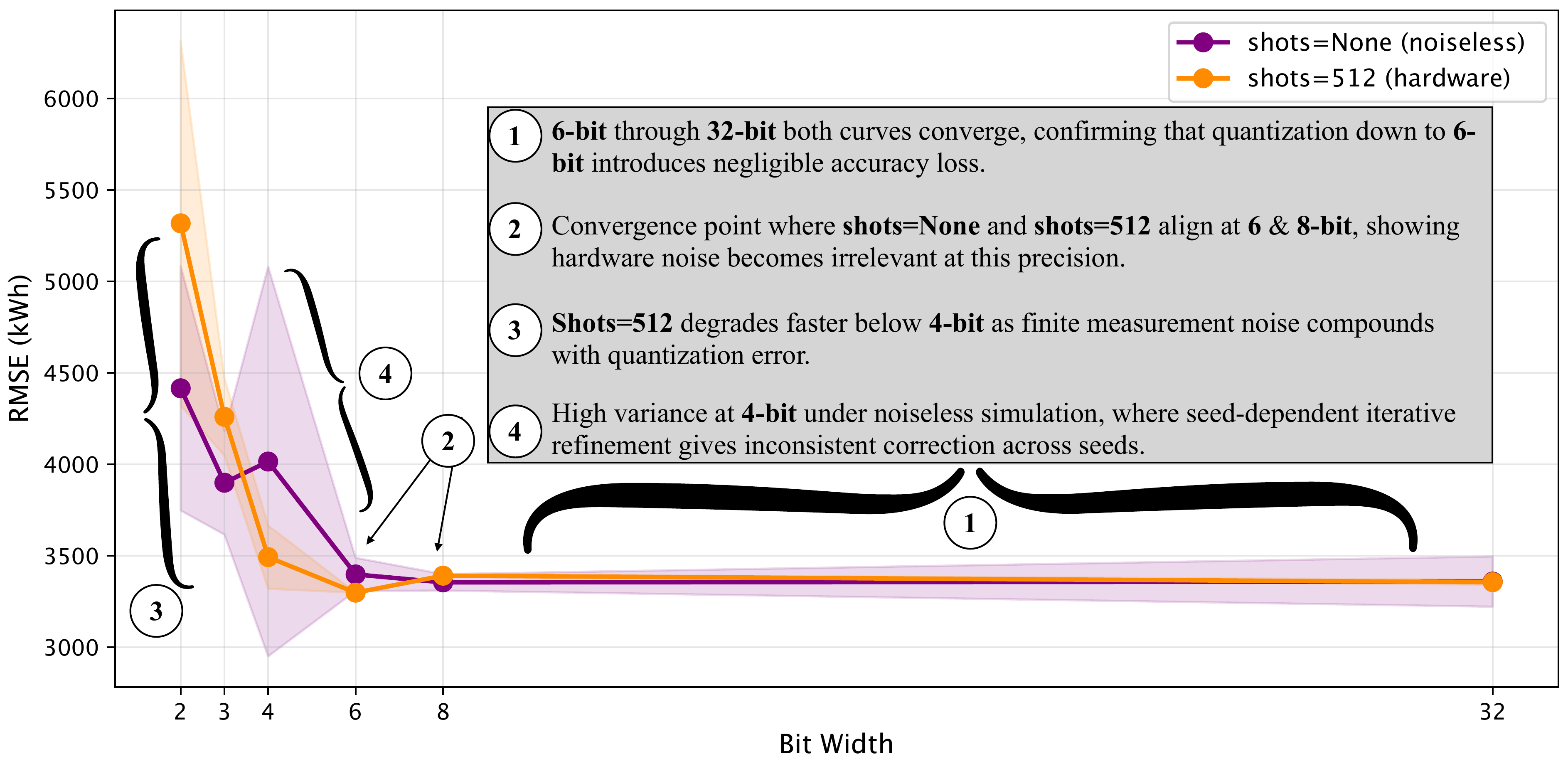}
 \caption{RMSE (kWh) versus bit width for quantized QRC readout under noiseless and finite-shot (\texttt{shots}=512) evaluation (mean $\pm$ std). Error is highest at very low precision, decreases markedly as bit width increases, and becomes closely aligned across both settings by 6 bits, remaining stable through 32 bits.}
    \label{p1}
\end{figure}

\begin{figure}[tpb]
    \centering
    \includegraphics[width=0.91\linewidth]{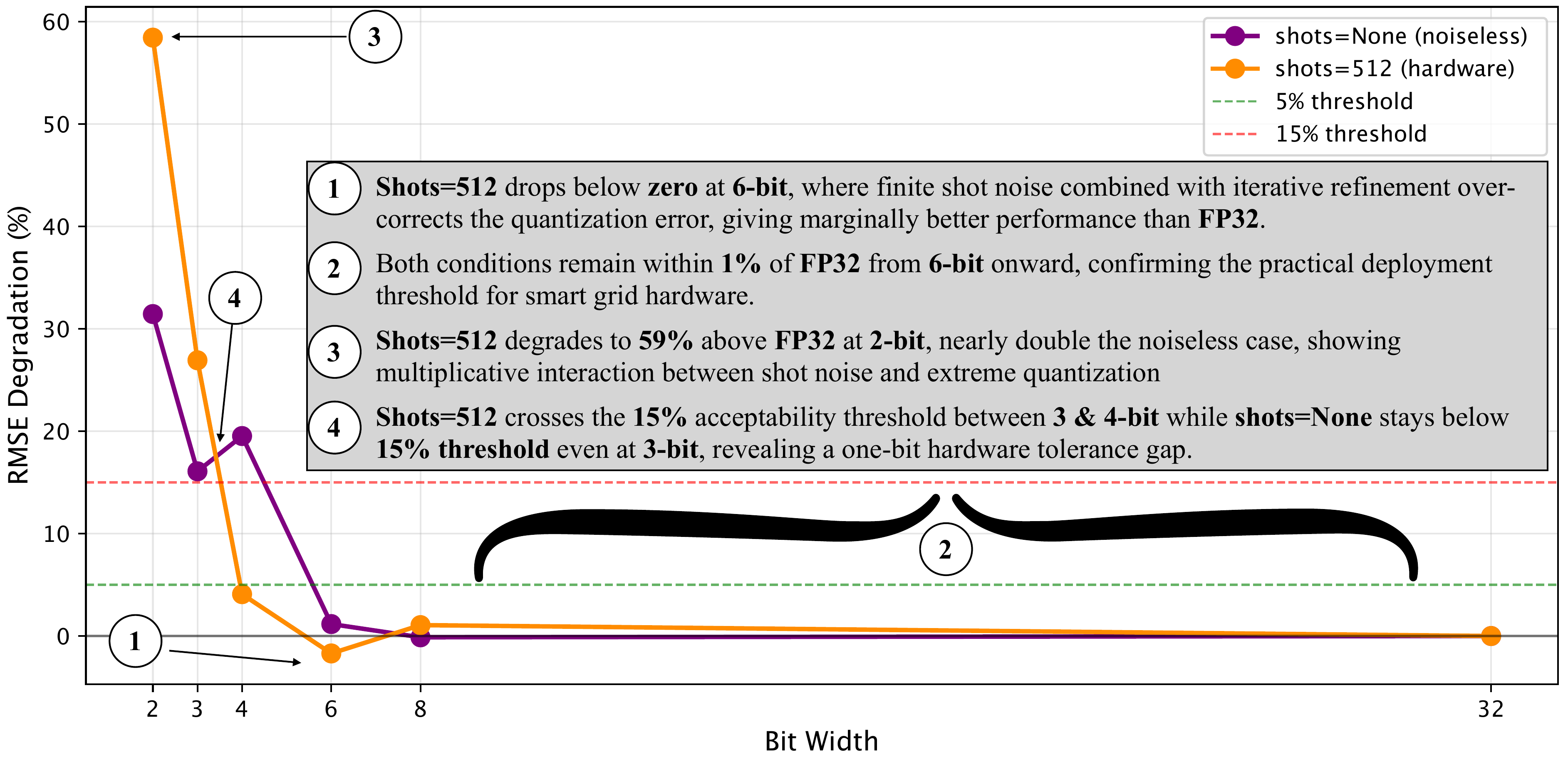}
    %\vspace{-0.8cm}
  \caption{Relative RMSE degradation (\% above the FP32 baseline) versus bit width under noiseless and finite-shot (\texttt{shots}=512) evaluation, with 5\% and 15\% reference thresholds. Error increases sharply at low precision, especially under finite-shot evaluation, while both curves fall below 5\% by 6 bits and approach zero at higher bit widths.}
    \label{p2}
\end{figure}

\begin{figure}[h]
    \centering
    \includegraphics[width=0.91\linewidth]{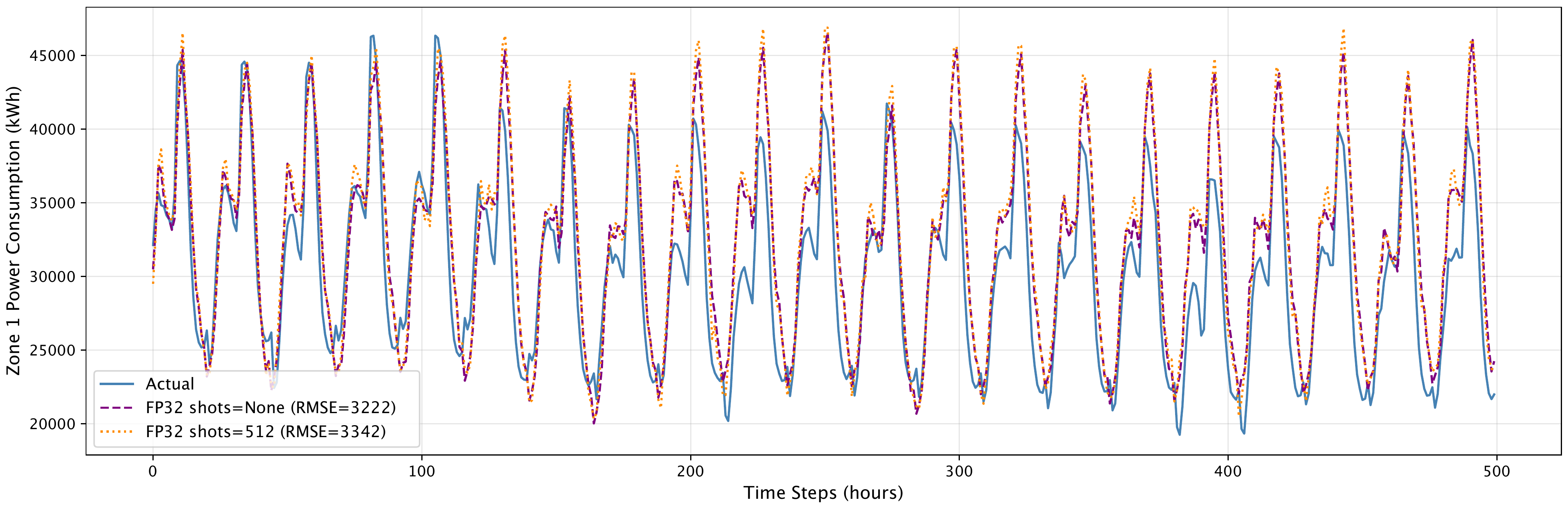}
    %\vspace{-0.8cm}
  \caption{Actual and predicted Zone 1 power consumption over 500 time steps for the selected QRC architecture. Both the noiseless (\texttt{shots=None}, RMSE = 3222) and finite-shot (\texttt{shots=512}, RMSE = 3342) FP32 predictions closely follow the observed signal, indicating that the reservoir captures the dominant temporal consumption pattern.}
    \label{p3}
\end{figure}

\vspace{-1pt}
Quantized QRC readout preserves forecasting accuracy at moderate precision while substantially reducing memory usage. RMSE decreases steadily as bit width increases, and the noiseless and finite-shot curves become closely aligned at 6-bit and above (Fig.~\ref{p1}), indicating limited additional error under shot-based evaluation in this range. 
The degradation trends in Fig.~\ref{p2} reinforce this observation: both settings fall below the 5\% threshold at 6-bit, whereas performance degrades more sharply below 4-bit, particularly in the finite-shot case. This behavior is also reflected qualitatively in Fig.~\ref{p3}, where the 6-bit quantized readout closely follows the FP32 prediction over 500 time steps. Table~\ref{tab:results} further shows that 6-bit achieves the best finite-shot RMSE of 3298.9 $\pm$ 0.3 while reducing readout memory by 81.2\% relative to FP32. Taken together, these results identify 6-bit as the most favorable compression-accuracy tradeoff observed in this study, with both 6-bit and 8-bit remaining within 1\% of full-precision performance.


\begin{table}[h]
\centering
\caption{RMSE (mean $\pm$ std) and memory savings across bit-widths 
under noiseless (shots=None) and finite-shot (shots=512) conditions.}
\label{tab:results}
\footnotesize
\begin{tabular}{cccc}
\hline
\textbf{Bit Width} & \textbf{shots=None} & \textbf{shots=512} & \textbf{Memory Saved} \\
\hline
32-bit & 3359.3 $\pm$ 137.2 & 3356.0 $\pm$ 13.6  & 0.0\%  \\
8-bit  & \textbf{3355.0} $\pm$ 45.1  & 3391.7 $\pm$ 6.7   & 75.0\% \\
6-bit  & 3398.5 $\pm$ 90.7  & \textbf{3298.9} $\pm$ 0.3   & 81.2\% \\
4-bit  & 4015.5 $\pm$ 1064.7 & 3493.2 $\pm$ 173.1 & 87.5\% \\
3-bit  & 3899.7 $\pm$ 283.6  & 4259.8 $\pm$ 211.8 & 90.6\% \\
2-bit  & 4415.8 $\pm$ 667.9  & 5317.5 $\pm$ 996.2 & 93.8\% \\
\hline
\end{tabular}
\end{table}
% =========================
% Conclusion
% =========================
\vspace{-3pt}
\section{Conclusion}
\vspace{-2pt}
In summary, 6-bit quantized QRC reduces readout memory by 81.2\% while maintaining accuracy within 1\% of the FP32 baseline under finite-shot evaluation. This identifies low-precision readout as a promising path toward deployment-efficient QRC for resource-constrained energy forecasting. The present study is limited to a small architecture search, two random seeds, one dataset, and finite-shot simulation rather than real quantum hardware. Future work will expand the search space, evaluate additional datasets, and explore hardware-aware quantization for real-device deployment.
%\vspace{-8pt}
\vspace{-2pt}
\section*{Acknowledgment}
\vspace{-2pt}
This work was supported in part by the NYUAD Center for Quantum and Topological Systems (CQTS), funded by Tamkeen under the NYUAD Research Institute grant CG008.
\end{spacing}
\vspace{-2pt}
\begin{spacing}{0.89}
\scriptsize
\bibliographystyle{ACM-Reference-Format}
\bibliography{refs}
\end{spacing}
\end{document}